\documentclass[pre,twocolumn,showpacs,preprintnumbers,amsmath,amssymb]{revtex4-1}

\usepackage[latin1]{inputenc}

\usepackage{graphics,dcolumn,bm,epic, eepic,fleqn,float}
\usepackage{graphicx}% Include figure files
\usepackage{dcolumn}% Align table columns on decimal point
\usepackage{bm}% bold math
\usepackage{epsfig}
\usepackage{color}

\usepackage{amssymb,amsmath,amsfonts,multirow,rotate,xcolor}
\usepackage{subfigure}

\newcommand{\lay}[1]{^{[#1]}}
\newcommand{\avg}[1]{\langle #1 \rangle}

\begin{document}

\title{Network structure of multivariate time series}

\author{Vincenzo Nicosia, Lucas Lacasa, and Vito Latora}
\affiliation{School of Mathematical Sciences, Queen Mary University of
  London, Mile End Road, E14NS London, UK}

\begin{abstract} 
Our understanding of a variety of phenomena in physics, biology and
economics crucially depends on the analysis of multivariate time
series. While a wide range of tools and techniques for time series
analysis already exist, the increasing availability of massive data
structures calls for new approaches for multidimensional signal
processing. We present here a non-parametric method to analyse
multivariate time series, based on the mapping of a multidimensional
time series into a multilayer network, which allows to extract
information on a high dimensional dynamical system through the
analysis of the structure of the associated multiplex network.  The
method is simple to implement, general, scalable, does not require
{\it ad hoc} phase space partitioning, and is thus suitable for the
analysis of large, heterogeneous and non-stationary time series.  We
show that simple structural descriptors of the associated multiplex
networks allow to extract and quantify nontrivial properties of
coupled chaotic maps, including the transition between different
dynamical phases and the onset of various types of synchronization. As
a concrete example we then study financial time series, showing that a
multiplex network analysis can efficiently discriminate crises from
periods of financial stability, where standard methods based on
time-series symbolization often fail.
\end{abstract}

\flushbottom \maketitle

\section{Introduction}

Time series analysis is a central topic in physics, as well as a
powerful method to characterize data in biology, medicine and
economics, and to understand their underlying dynamical origin. In the
last decades, the topic has received input from different disciplines
such as nonlinear dynamics, statistical physics, computer science or
Bayesian statistics and, as a result, new approaches like nonlinear
time series analysis~\cite{nonlinear} or data mining~\cite{data} have
emerged. More recently, the science of complex networks
\cite{barabasirev,Boccaletti2006, Newman2010} has fostered the growth
of a novel approach to time series analysis based on the
transformation of a time series into a network according to some
specified mapping algorithm, and on the subsequent extraction of
information about the time series through the analysis of the derived
network.  Within this approach, a classical possibility is to
interpret the interdependencies between time series (encapsulated for
instance in cross-correlation matrices) as the weighted edges of a
graph whose nodes label each time series, yielding so called
functional networks, that have been used fruitfully and extensively in
different fields such as neuroscience \cite{Bullmore2009} or finance
\cite{financenet}. A more recent perspective deals with mapping the
particular structure of univariate time series into abstract graphs
\cite{Small, Thurner2007, Small2, donner2010, donner2011, pnas,
  seminalPRE}, with the aims of describing not the correlation between
different series, but the overall structure of isolated time series,
in purely graph-theoretical terms. Among these latter approaches, the
so called visibility algorithms \cite{pnas, seminalPRE} have been
shown to be simple, computationally efficient and analytically
tractable methods \cite{analytic, Gutin}, able to extract nontrivial
information about the original signal \cite{EPL}, classify different
dynamical origins \cite{Toral} and provide a clean description of low
dimensional dynamics \cite{chaos, quasi, intermitencia}. As a
consequence, this particular methodology has been used in different
domains including earth and planetary sciences \cite{geo, geo2,
  astro}, finance \cite{finance} or biomedical fields \cite{bio} (see
\cite{review} for a review). Despite their success, the range of
applicability of visibility methods has been so far limited to
univariate time series, whereas the most challenging problems in
several areas of nonlinear science concern systems governed by a large
number of degrees of freedom, whose evolution is indeed described by
multivariate time series.

\begin{figure}[!tp]
  \begin{center}
    \includegraphics[width=0.9\columnwidth]{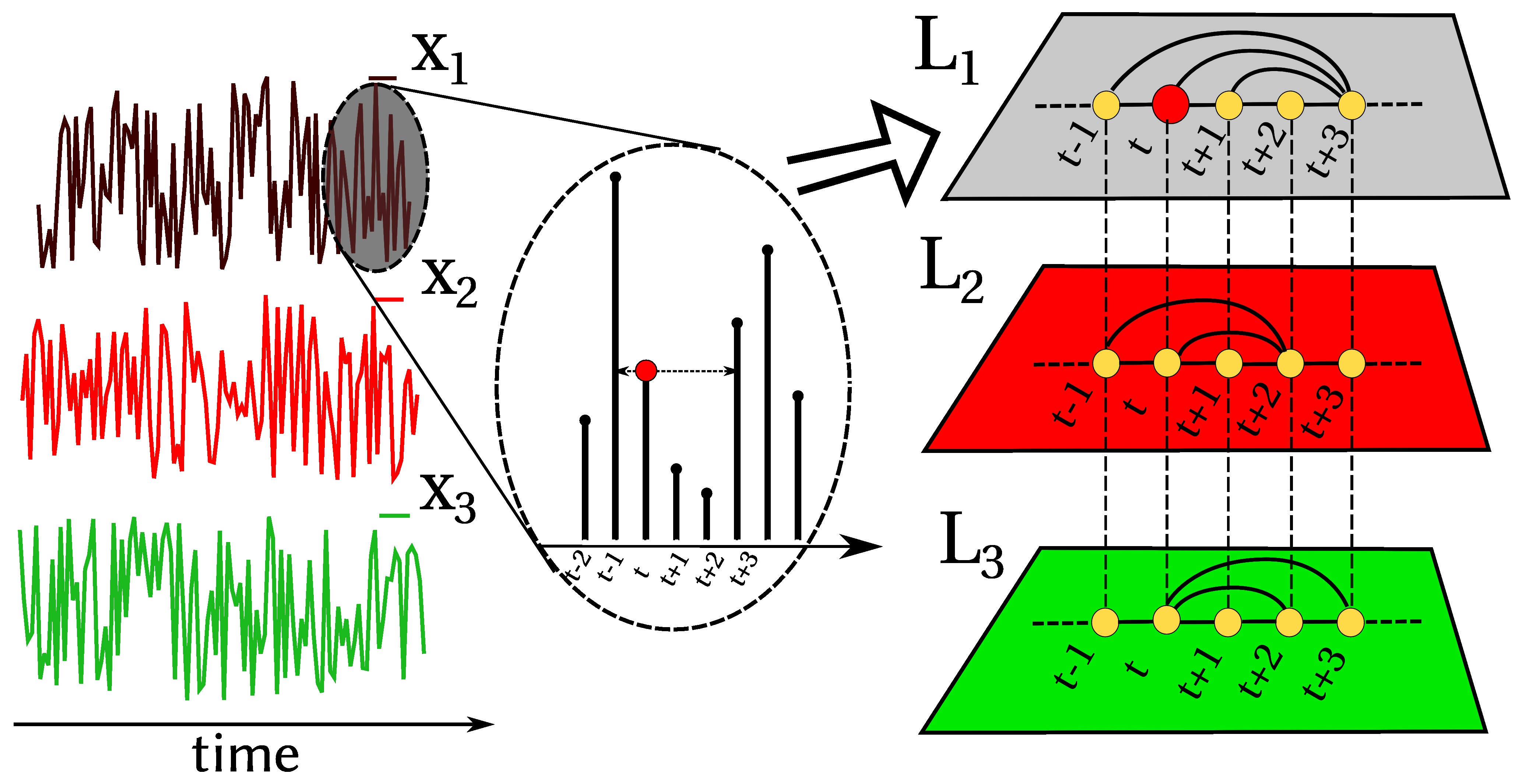}
  \end{center}
  \caption{(color online) The Horizontal Visibility Graph (HVG)
    algorithm maps a \textit{M}-dimensional time series
    $\{x^{[\alpha]}(t)\}_{t=1}^N, \alpha=1,\ldots,M$, into a 
    {\em multiplex visibility graph} $\cal M$, i.e. a multi-layer network
    where each layer $\alpha$ is the HVG of the $\alpha$-th component 
    of the time series.}
  \label{fig:fig0}
\end{figure}
\begin{figure*}[!t]
\centering \includegraphics[width=6.2in]{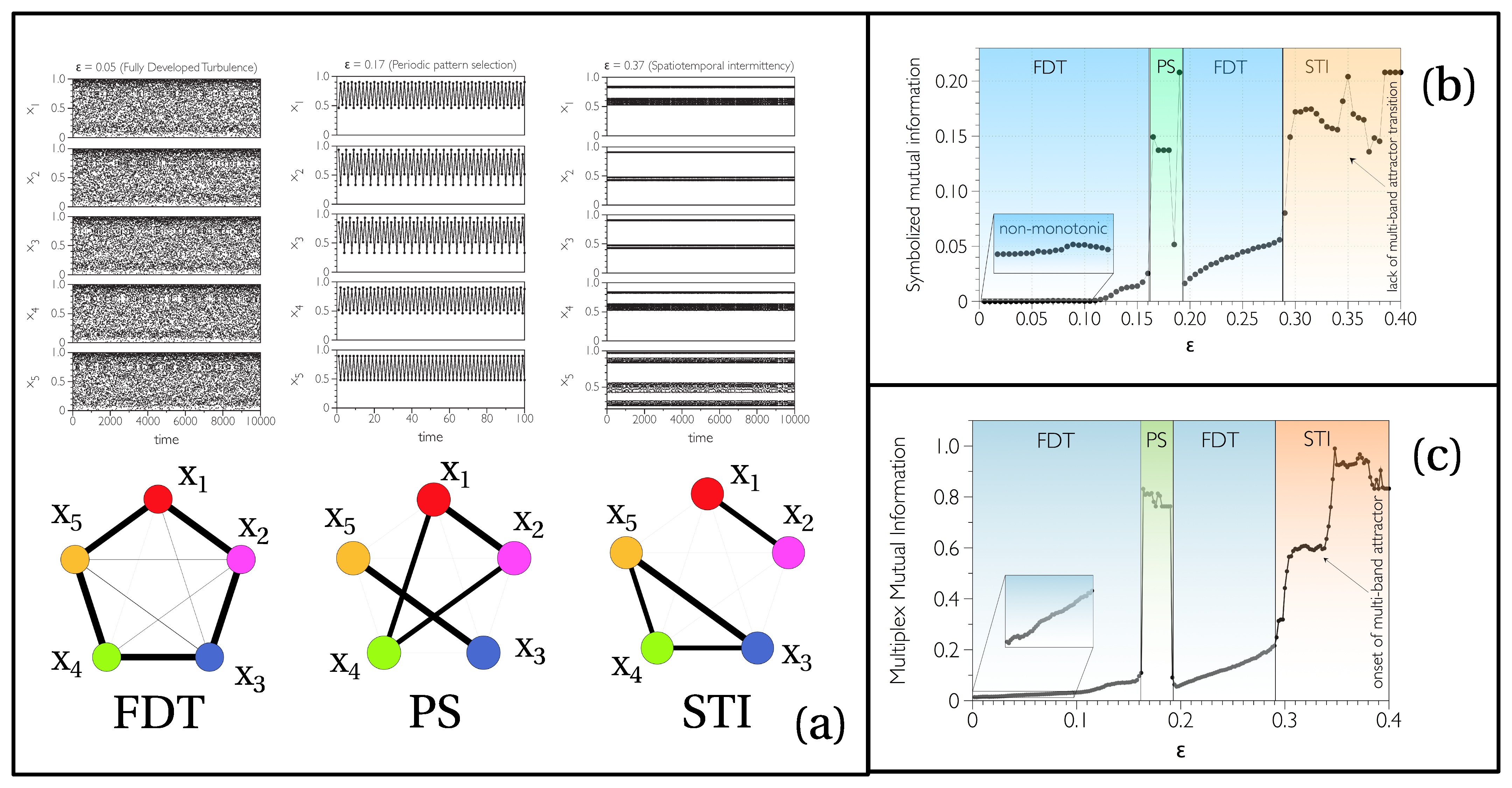}
\caption{(color online) (a) Sample time series generated by five
  diffusively coupled fully chaotic logistic maps at different values
  of the coupling strength $\epsilon$, showing respectively Fully
  Developed Turbulence (FDT) at $\epsilon=0.05$, Pattern Selection
  (PS) at $\epsilon= 0.17$, and Spatio-temporal Intermittency (STI) at
  $\epsilon= 0.37$. Information flow among units is well captured by
  projecting the multiplex visibility graph $\cal M$ into a layer
  graph $\cal G$ (bottom), whose nodes represent layers and the edges
  weights (their thickness in Figure) the mutual information among
  them. (b) The average pairwise mutual information $I^{\text{SYMB}}$
  computed from pre-symbolized time series, or (c) the corresponding 
  version $I$ computed from the associated multiplex
  visibility graph can both be used as order parameters to distinguish
  different dynamical phases (see SI for additional analysis with
  average edge overlap). However, only the multiplex
  measure is capable of detecting fine-grain structures such as the
  onset of multi-band chaotic attractors.}
\label{comparison}
\end{figure*} 

In order to fill this gap, in this work we introduce a visibility
approach to analyze multivariate time series based on the mapping of a
multidimensional signal into an appropriately defined multi-layer
network~\cite{Bianconi2013,Nicosia2013,dedomenico,kivela2014,Boccaletti2014,Battiston2014},
which we call \textit{multiplex visibility graph}.  
Taking advantage of the recent development in the theory of multilayer networks 
\cite{dedomenico,kivela2014,Boccaletti2014,Battiston2014,Bianconi2013,
  NicosiaNonLinear, NicosiaCorrelations}, new information can be
extracted from the original multivariate time series, with the aims of
describing signals in graph-theoretical terms or to construct novel
feature vectors to feed automatic classifiers in a simple, accurate
and computationally efficient way. We will show that, among other
possibilities, a particular projection of this multilayer network produces 
a (single-layer) network similar in spirit to functional
networks, while being more accurate than standard methodologies to
construct these. We validate our method by investigating the rich 
high-dimensional dynamics displayed by canonical models of spatio-temporal
chaos, and then apply our framework to describe and quantify periods
of financial instability from a set of empirical multivariate
financial time series.

\section{Results}

Let us start by recalling that visibility algorithms are a family of
geometric criteria which define different ways of mapping an ordered
series, for instance a temporal series of $N$ real-valued data
$\{x(t)\}_{t=1}^N$, into a graph of $N$ nodes. The standard linking
criteria are the natural visibility \cite{pnas} (a convexity
criterion) and the horizontal visibility \cite{seminalPRE} (an
ordering criterion). In the latter version, two nodes $i$ and $j$ are
linked by an edge if the associated data $x(i)$ and $x(j)$ have
horizontal visibility, i.e. if every intermediate datum $x(k)$
satisfies the ordering relation $x(k) <\inf\{x(i),x(j)\}, \ \forall k:
\ i < k < j.$ The resulting Horizontal Visibility Graph (HVG) is an
outerplanar graph (indeed a subgraph of the original visibility
graph), whose topological properties have been shown to be
analytically tractable for a large class of different dynamical
processes \cite{seminalPRE, analytic}. Both the natural and horizontal
graphs are undirected, however directed graphs can be easily
constructed by by distinguishing ingoing from outgoing links with
respect to the arrow of time, something which has proven useful to
assess time asymmetries and to quantify time series
irreversibility~\cite{irrev, irrev2}.

Consider a $M$-dimensional real valued time series
$\{\textbf{x}(t)\}_{t=1}^N$, with $\textbf{x}(t) = (x\lay{1}(t),
x\lay{2}(t),\ldots, x\lay{M}(t))\in\mathbb{R}^M$ for any value of $t$,
measured empirically or extracted from a $M$-dimensional, either
deterministic or stochastic, dynamical system.
%$\textbf{x}(t+1)=\textbf{F}\left(\textbf{x}(t)\right)$
An $M$-layer multiplex network, that we call the {\em multiplex
  visibility graph} $\cal M$ is then constructed, where layer $\alpha$
corresponds to the HVG associated to the time series of state variable
$\{ x\lay{\alpha}(t) \}_{t=1}^N$. We illustrate this procedure for
$M=3$ in Fig.~\ref{fig:fig0}.  Note that in this work we focus on the
undirected, horizontal version, but other visibility linking criteria
can be analogously used to define different multiplex versions.
 $\cal M$ is represented by the vector of adjacency matrices of its
layers $\mathcal{A}=\{A\lay{1}, A\lay{2}, \ldots, A\lay{M}\}$, where
$A\lay{\alpha} = \{a\lay{\alpha}_{ij}\}$ and $a\lay{\alpha}_{ij}=1$ if
and only if node $i$ and node $j$ are connected by a link at layer
$\alpha$ \cite{Battiston2014,Nicosia2013}.  
This builds a bridge between multivariate series analysis and the
recent developments in the study of multilayer networks 
\cite{dedomenico,kivela2014,Boccaletti2014,Battiston2014,Bianconi2013,NicosiaNonLinear, NicosiaCorrelations}. Among the different 
possible multiplex measures that with our mapping 
is now possible to exploit also 
in the context of multidimensional time series, 
we focus here on one which allows to detect and quantify 
inter-layer degree correlations~\cite{NicosiaCorrelations}. 
In such a way we can characterize information shared 
across variables (layers) of the
underlying high dimensional system, 
this aspect being
indeed of capital importance in fields such as neuroscience or 
economics and finance. Given a pair of layers $\alpha$ and $\beta$ of
$\cal M$, respectively characterized by the degree distributions
$P(k\lay{\alpha})$ and $P(k\lay{\beta})$, we can define an {\em
  interlayer mutual information} $I_{\alpha,\beta}$ as:
\begin{equation}
  I_{\alpha,\beta}=\sum_{k\lay{\alpha}}\sum_{k\lay{\beta}}P(k\lay{\alpha},
  k\lay{\beta}) \log{\frac{P(k\lay{\alpha},
      k\lay{\beta})}{P(k\lay{\alpha})P(k\lay{\beta})}}
  \label{eq:MI}
\end{equation}
where $P(k\lay{\alpha}, k\lay{\beta})$ is the joint probability to
find a node having degree $k\lay{\alpha}$ at layer $\alpha$ and degree
$k\lay{\beta}$ at layer $\beta$. In general, the higher $I_{\alpha,
  \beta}$ the more correlated the degree distributions of the two
layers and, therefore, the structure of the associated time
series. If we then average the quantity $I_{\alpha,\beta}$ over every
possible pair of layers of $\cal M$, we obtain a scalar variable
$I=\avg{I_{\alpha, \beta}}_{\alpha, \beta}$ which captures the typical 
amount of information flow in the system.
Of course, the values $\{I_{\alpha,\beta}\}$ can be considered as the
weights of the edges of a \textit{graph of layers} $\mathcal{G}$,
this being a projection of the original multiplex
visibility graph $\mathcal{M}$ into a (single-layer) weighted graph of
$M$ nodes, where each node represents one layer. Edges in this 
case denote mutual information, but our approach can be 
easily generalized so that edges can 
represent different types of interdependence, such as
correlation~\cite{Bullmore2009}, causality~\cite{grangernetwork},
etc.~\cite{Sporns2013} between layers. Hence for this particular
projection, one is actually analyzing the visibility analog of standard 
functional networks. We shall indeed see that the multiplex 
projection is genuinely different and often works better than other 
ways to construct functional networks. This is mainly because the latters 
often require, to
compute the weight of each edge, the symbolization of each time series, and
this pre-processing is usually afflicted by several limitations and
ambiguities \cite{nonlinear}. As we show in the online Supplementary
Information (SI), the technique proposed here also appears to be free
from these well known ambiguities.\\

\begin{figure*}[!t]
  \begin{center}
    \includegraphics[width=6.2in]{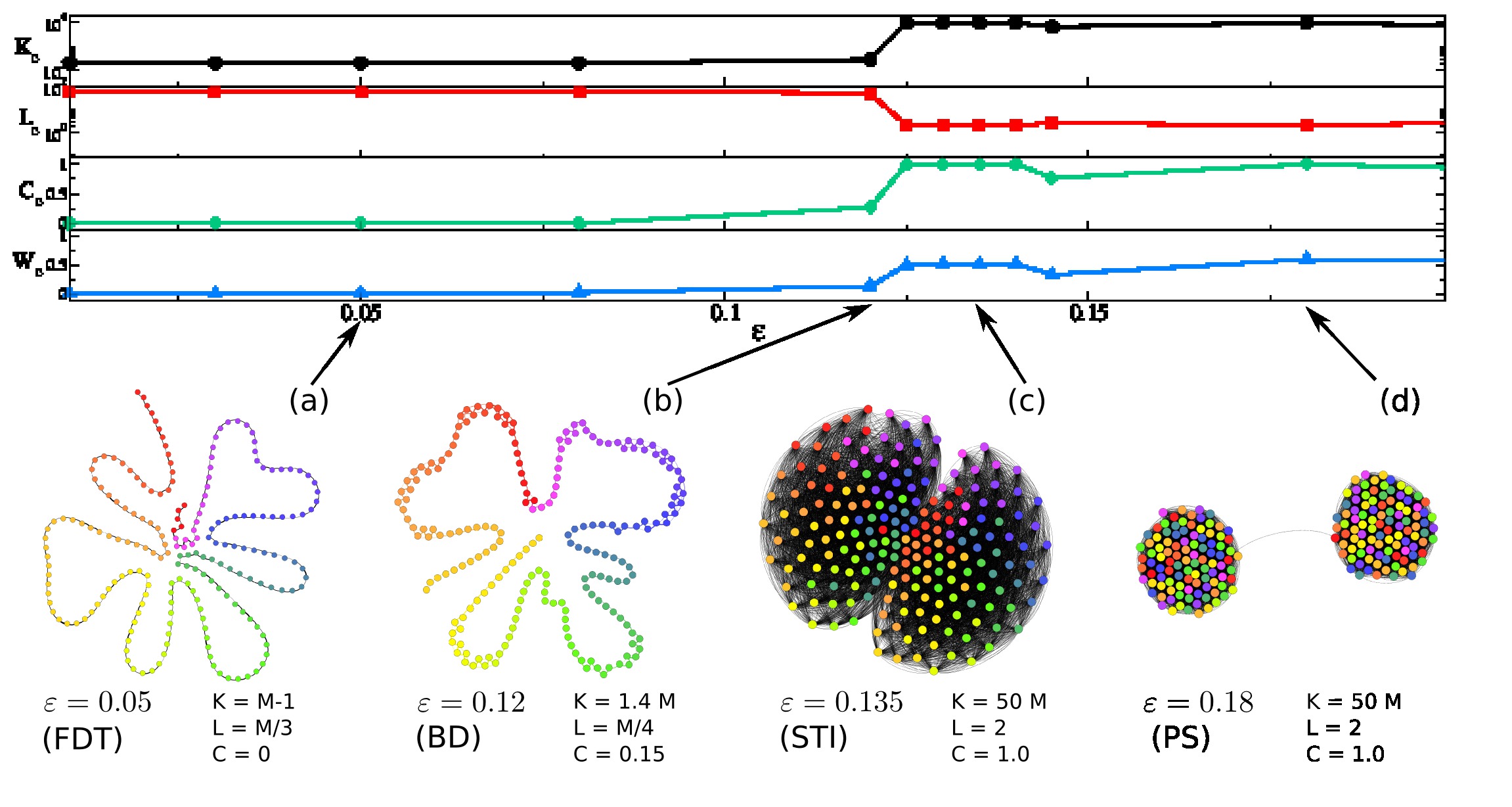}
  \end{center}
  \caption{(color online) Backbone graphs $\mathcal{G}'$ of the graph
    of layers $\cal G$ obtained from the multiplex visibility graph of
    $M=200$ diffusively coupled chaotic maps for different values of
    $\epsilon$. For each dynamical phase, $\mathcal{G}'$ has a
    different structure. In FDT, $\mathcal{G}'$ is a chain, revealing
    the diffusive nature of the coupling in this weakly interacting
    situation. In PS, $\mathcal{G}'$ is formed by non-interacting
    communities, whereas in STI -where the ghost of the periodic
    attractor is perturbed by chaotic excursions-, $\mathcal{G}'$ is
    formed of slightly overlapping dense communities. We also report
    several topological properties of $\mathcal{G}'$ (number of edges
    $K$, average shortest path length $L$, clustering coefficient $C$,
    and total weight of the edges) that have different qualitative
    values in each phase. Nodes colours determine the map label, so
    similar colours denote maps at close distance in the CML.}
  \label{fig:fig3}
\end{figure*}

\subsection{Information flow and phase diagram in Coupled Map Lattices.} As a case
study we first consider diffusively coupled map lattices (CMLs),
high-dimensional dynamical systems with discrete time and continuous
state variables, widely used to model complex spatio-temporal dynamics
~\cite{kaneko} in as disparate contexts as turbulence
\cite{kanekobook}, or vacuum fluctuations and dark energy
\cite{beck}. Namely we consider a ring of $M$ sites, and we assume
that the dynamical evolution of the state $x\lay{\alpha}$ of site
$\alpha$ is determined by:
\begin{align}
x\lay{\alpha}(t+1) =(1-\epsilon)f[x\lay{\alpha}(t)] + 
 \frac{\epsilon}{2}\bigg(f[x\lay{\alpha-1}(t)]+f[x\lay{\alpha+1}(t)]\bigg),
\label{CML}
\end{align}
$\forall \alpha=1,\dots, M$, where $\epsilon \in [0,1]$ is the
coupling strength, and $f(x)$ is typically a chaotic map. For
different values of $\epsilon$ and $M$ CMLs display a very rich phase
diagram, which includes different degrees of synchronization and
dynamical phases such as Fully Developed Turbulence (FDT, a phase with
incoherent spatiotemporal chaos and high dimensional attractor),
Pattern Selection (PS, a sharp suppression of chaos in favor of a
randomly selected periodic attractor), or different forms of
spatio-temporal intermittency (STI, a chaotic phase with low
dimensional attractor). The origin of such a rich and intertwined
structure comes from the interplay between the local tendency towards
inhomogeneity, induced by the chaotic dynamics, and the global
tendency towards homogeneity in space, induced by the diffusive
coupling.  Fig.~\ref{comparison} report the results obtained for a CML
of $M=5$ diffusively coupled, fully chaotic logistic maps
$f(x)=4x(1-x)$, which exhibits several transitions from high
dimensional chaos, to pattern selection, to several forms of partially
synchronized chaotic states when $\epsilon$ is increased.  The plots
of Fig.~\ref{comparison} are based on averages over 100 realisations
of the CML dynamics. For each realisation, we constructed a
multivariate time series
$\{\textbf{x}_1,\textbf{x}_2,...,\textbf{x}_N\}$ of $N=2^{14}$ data
points (discarding the transient) and we generated the corresponding
multiplex visibility graph. In Fig.~\ref{comparison}(b) we show how
the average mutual information $I$ of the multiplex visibility graph
associated to the system (see SI for other multiplex measures) is able
to distinguish between the different phases.  In particular, $I$ is a
monotonically increasing function of $\epsilon$ in the FDT phases, and
therefore quantifies the amount of information flow among
units. Notably, it also detects qualitative changes in the underlying
dynamics (such as the chaos suppression in favor of a randomly
selected periodic pattern, or the onset of a multi-band chaotic
attractor during intermittency) and therefore can be used as a scalar
order parameter of the system. For comparison, in
Fig.\ref{comparison}(c) we also plot the corresponding quantity
derived from a standard functional network analysis. Namely,
$I^{\text{SYMB}}$ is the average mutual information computed directly
on the multivariate time series, after performing the necessary time
series symbolization. Although there are qualitatively similarities,
subtle aspects such as the monotonic increase of synchronization with
$\epsilon$ in FDT, or the onset of multiband attractors in STI are not
captured by $I^{\text{SYMB}}$ (additional details comparing our method
to standard functional network approaches can be found in the SI). In
panel (a) of the same Figure we also report the projections of the
three multiplex networks $\cal M$ into the corresponding graphs of
layers $\cal G$, whose edge widths are proportional to the values
$I_{\alpha, \beta}$ of mutual information between layers $\alpha$ and
$\beta$.  A simple visual inspection of such graphs reveals the
different type of information flow among units, depending on the
dynamical phases of the system.  In particular, notice that the
diffusive nature of the coupling emerges in the ring-like structure of
graph $\cal G$ corresponding to weakly interacting maps (FDT) (the
analysis is extended in SI to globally coupled maps, these being a
mean-field version of CML where complete synchronisation is possible,
showing that our method correctly detects the onset of this new
regime).

\subsection{Scaling up the system.} The previous study suggests that
the quantities $\{I_{\alpha, \beta}\}$ (see Fig.~\ref{comparison}(a))
accurately capture relevant information of the underlying dynamics. To
further explore this aspect and to assess scalability, we considered a
chain of $M=200$ diffusively coupled logistic maps, each governed by
Eq.~(\ref{CML}). New short dynamical phases, such as the so called
Brownian motion of Defects (BD) --a transient phase between FDT and
PS--, emerge when the dimension of the system is increased, and as the
description gets more cumbersome, projections and coarse-grained
variables are needed~\cite{kaneko}.  Since the graph of layers
$\mathcal{G}$ is by construction a complete graph (just as any
functional network), for visual reasons in Fig.~\ref{fig:fig3} we
report the structural properties of $\mathcal{G}'$, the backbone of
$\mathcal{G}$ obtained starting from an empty graph of $M$ nodes and
adding edges sequentially in decreasing order of $I_{\alpha,\beta}$,
until the resulting graph consists of a single connected
component. The structure of $\mathcal{G}'$ is unique for each phase
and qualitatively different across phases, thus providing a simple
qualitative way to portrait different dynamics in high-dimensional
systems.\\

\begin{figure*}[!t]
  \begin{center}
    \includegraphics[width=6.2in]{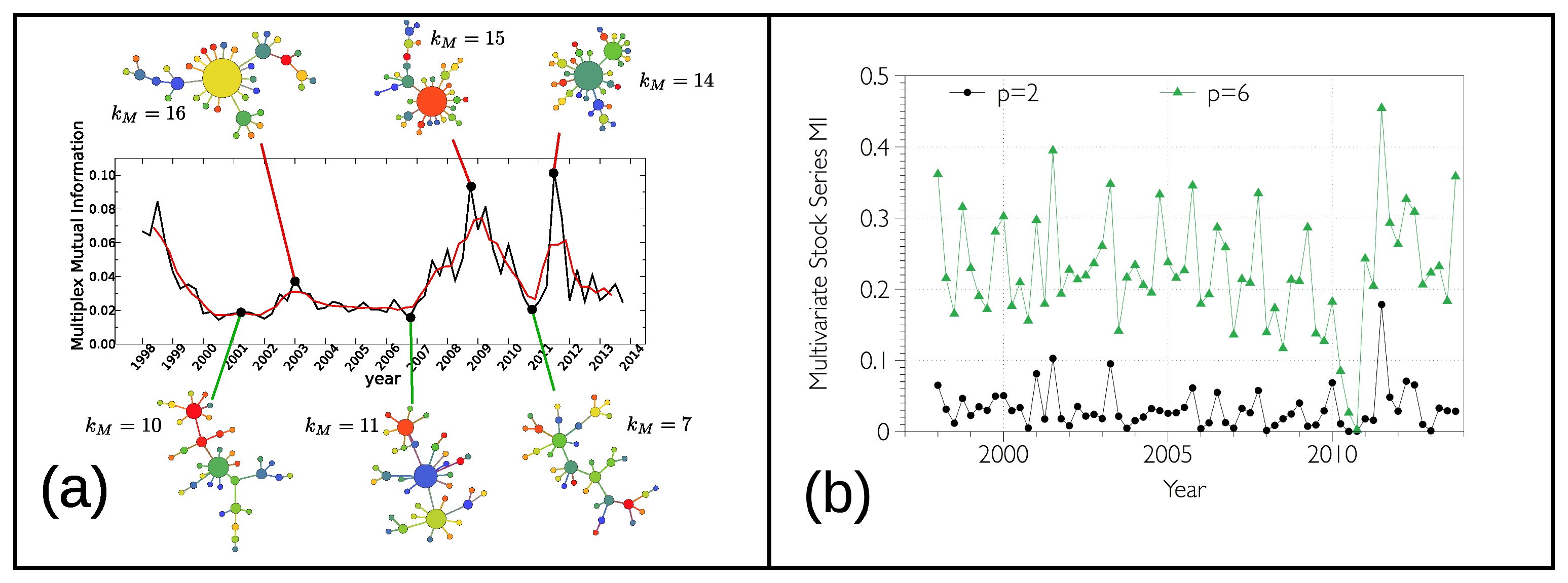}
  \end{center}
  \caption{(a) The multiplex mutual information is a suitable
    quantity to detect global changes of behavior in
    multivariate financial time series. In the plot we report the
    value of the average information (the red line is the corresponding 
    running average) computed from the multiplex networks
    constructed from price time series of 35 major assets in NYSE and NASDAQ
    in each 3-month period between January 1998 and December 2013
    (see Methods for details).  Notice that the most pronounced peaks of mutual
    information correspond to periods of major
    financial instability (the .com bubble in 1998-1999, the mortgage
    subprime crisis in 2007-2012). The Maximum Spanning Trees of the 
    corresponding networks of layers (six
    typical examples are shown), always include a large
    hub during crises (the three top networks), whose degree
    $k_{M}$ is larger than the maximum degree observed in periods of 
    stability (the three bottom networks). Each asset is assigned 
    the same color in all the networks, while the size 
    of a node is proportional to its degree. (b) The mutual
    information among the same set of time series performed after a
    standard symbolization is not able to single out crises. The
    resulting signal is much more herratic and not as informative as
    the multiplex mutual information shown in panel (a). }
\label{MIseries_refined}
\end{figure*}
    
\subsection{Multiplex analysis of financial instabilities.} 
As an example of the possible applications of the multiplex visibility
graph approach to the analysis of real-world multivariate time
series, we report a study of the prices of financial assets. 
Namely, we considered the time evolution of 
stock prices of $M=35$ of the largest US companies by market capitalization
from NYSE and NASDAQ (see SI for details) over the period
1998-2013. 
The $M$ time series have a very high resolution (one data per minute),
yielding $O(2\cdot10^6)$ data per company.  We divided each
multivariate time series into non-overlapping periods of three months
(quarters), and we constructed a {\it temporal} multiplex visibility
graph consisting of 64 multilayer snapshots, each formed by the
35-layer multiplex visibility graph corresponding to one of the 
3-months periods. We then investigated the time evolution of the 
multiplex mutual information among layers, and  how this correlates 
to the presence of periods of financial instability.
  
In Fig.~\ref{MIseries_refined}(a) we plot the value of the average 
multiplex mutual information $I$ as a function of time. For comparison we also
report in Fig.~\ref{MIseries_refined}(b) the analogous measure
computed directly on the original series, after an appropriate
symbolization (see SI for details). We find that the multiplex visibility 
graph approach captures the onset
of the major periods of financial instability (1998-1999,
corresponding to the .com bubble, and 2007-2012, corresponding to the
great recession that took place as a consequence of the mortgage
subprime crisis), which are characterised by a relatively increased 
synchronisation of stock prices, clearly distinguishing them from the
seemingly unsynchronised interval 2001-2007, which in turn corresponds
to a more stable period of the economy.  In direct analogy with the language used for 
CMLs, we could say that in periods of financial stability, the
system is close to equilibrium, degrees of freedom are evolving in a
quasi-independent way, reaching a fully developed turbulent state of 
low mutual information (hence unpredictable and efficient from a financial
viewpoint ). On the other hand, during periods of financial instability 
(bubbles and crisis) the system is externally perturbed, hence driven
away from equilibrium, and the degrees of freedom share larger mutual
information (the system is therefore less unpredictable and inefficient from a
financial viewpoint). 
As shown in Figure \ref{MIseries_refined}(b), an  
analogous analysis based on the symbolization of the time series fails 
to capture all such details (see SI for additional analysis).
Finally, as also seen in the case of
the multiplex visibility graphs associated to CMLs, the differences in the
values of average mutual information corresponding to different phases
are indeed related to a different underlying structure of the network
of layers. In Fig.\ref{MIseries_refined} we show the Maximal Spanning
Trees (MST) of the networks of layers associated to six typical time
windows~\cite{Mantegna1999,financenet}. The three networks at the
bottom of the Figure represents periods of financial stability, 
while those at the top of the Figure correspond to the three local 
maxima of mutual information. Interestingly, the MSTs  
in periods of financial instability all have a massive hub which is
directly linked to as much as $50\%$ of all the other
nodes. Conversely, the degree is more evenly distributed in the MSTs
associated to periods of economic stability.

\section{Discussion}
The approach based on multiplex visibility graphs introduced in this 
work provides an alternative and powerful 
method to analyze multivariate time series. We have first validated 
our method focusing on signals whose underlying dynamics
is well known and showing that measures describing the structure 
of the corresponding multiplex networks (which are not
affected by the usual problems of standard symbolization
procedures)  are able to capture and quantify the
onset of dynamical phases in high-dimensional coupled chaotic maps, as
well as the increase or decrease of mutual information among layers
(maps) within each phase. We then have studied an application to the
analysis of multivariate financial series, showing that 
multiplex measures, differently from other standard methods, 
can easily distinguish periods of financial stability 
from crises, and can thus be used effectively as a support tool in
decision making. 
\\
The proposed method is extremely flexible and can be used in all
situations where the dynamics is poorly understood or unknown, with
potential applications ranging from fluid dynamics to neuroscience or
social sciences. In this article we have focused only on a particular
aspect, which is the characterization of the information flow among
the different variables of the system,  
and we have consequently based our analysis on the
study of the resulting networks of layers. However, our approach is
quite general, and the mapping of multivariate time series into
multiplex visibility graphs paves the way to the study of the
relationship between specific structural descriptors recently
introduced in the context of multiplex networks and the properties of
real-world dynamical systems. We are confident that our method is 
only the first step towards the construction of feature-based 
automatic tools to classify dynamical systems of any kind.

%%%\bibliography{apssamp}% Produces the bibliography via BibTeX.

\section*{Acknowledgments}
  V.N. and V.L. acknowledge support from the Project LASAGNE, Contract
  No.318132 (STREP), funded by the European Commission. V.L. acknowledges  
  support from the EPSRC project GALE, EP/K020633/1.

\section*{Author contributions statement}

All the authors conceived the study, performed the experiments,
analysed the results and wrote the paper. All the authors approved the
final version of the manuscript.

\end{document}